# Logic Programming for an Introductory Computer Science Course for High School Students

Timothy Yuen[1], Maritz Reyes[2], Yuanlin Zhang[3]

[1] The University of Texas at San Antonio, USA
[2] The University of Texas at Austin, USA
[3] Texas Tech University, USA
Timothy.Yuen@utsa.edu, maritza_reyes@utexas.edu, y.zhang@ttu.edu

**Abstract.** This paper investigates how high school students approach computing through an introductory computer science course situated in the Logic Programming (LP) paradigm. This study shows how novice students operate within the LP paradigm while engaging in foundational computing concepts and skills, and presents a case for LP as a viable paradigm choice for introductory CS courses.

## 1 Introduction

The debate in paradigm selection in CS0 and CS1 type courses is often split between object-oriented (OO) and procedural paradigms, which also leads to discussions on programming language choice. Selection is a difficult task as there are no standard languages or paradigms to use in the field. Though both these approaches have been successful in introductory CS courses [1], some research has also shown only minimal differences when comparing the outcomes between paradigms [2, 3]. At the same time, research on the teaching of introductory CS courses also reveals the limitations of those approaches [4, 5] (e.g., the suitability of Java, C, C++ for education is debatable [4] as well as the objects-first). However, the Logic Programming (LP) based approach is largely ignored by the CS0/CS1 community [4] despite continuing efforts in both teaching and research on LP.

### 1.1 Logic Programming and CS Education

Prolog may be the most well-known instance in which LP was used to teach CS. In particular, Prolog was used to teach children in the 1980s [6, 7]. As a pioneer, Kowalski [6] focused on the declarative aspects of Prolog and restricted the use of procedural aspects of Prolog to a minimum. Later, researchers and practitioners found that the procedural aspects of Prolog have been the main source of misconceptions and difficulties [8], while the benefits of its declarative aspects were acknowledged. In the last two decades, Prolog has made some appearances: listed in high school curriculum [9], and taught in gifted and talented high school students [10] and in general high school/undergraduate students [11], though it does not enjoy the same attention as procedural and OO paradigms.

### 1.2 Advantages of Logic Programming

From the existing work and practice in LP and general programming [4, 5, 9, 12], the main advantages of LP as related to computing education are: simple syntax and intuitive/declarative semantics, natural connection to abstraction, logic reasoning and knowledge representation which form the foundation of computing and other disciplines, involvement of non-trivial, interesting and challenging problems (e.g., Sudoku problem) early, and the mathematical flavor of LP. LP also provides a rich context for embedded essential computing concepts and skills. As for industrial relevance, LP and Declarative Programming (DP), in general, have seen their profound application and impact in database query languages, problem (formal) specification languages, and domain specification languages including popular web application development languages such as HTML, CSS and XSLT etc. A breakthrough in LP research in the last two decades is the establishment of the Answer Set Programming (ASP) [13, 14] paradigm which has inherited the declarative nature of Prolog while fully getting rid of its procedural features. Currently, ASP is a dominating LP formalism in knowledge representation.

### 1.3 Research Questions

Although there are some movements [e.g., 15] in expanding LP, research on how ASP, with full declarative nature, facilitates student learning in introductory computer science is still absent. The research question for this study is: *How does adopting ASP to teach an introductory CS course for high school students impact their understanding of computer science and computing?*

## 2 Methods

This study adopts a qualitative approach in exploring, in-depth, how high students' understanding of CS and computing through LP, specifically ASP.

### 2.1 Setting and Participants

Participants were recruited from the TexPREP program at Texas Tech University during Summer 2015. TexPREP is a summer program for 6$^{th}$-12$^{th}$ graders and offers full day courses in science, engineering, math, and computer science over several weeks. Courses are typically taught by local K-12$^{th}$ grade teachers as well as university faculty members and students. In the case of this study, the class was taught by the third author with some TAs (the second author was also a TA). TexPREP is a selective program:

students had to have an A/B+ average, letters of recommendations, a written, and a personal interview for consideration. Prospective participants were in their fourth year of TexPREP taking Computer Science 4 (CS4). They have already taken its CS2 (Scratch [17]) and CS3 (Alice [18]) courses in the CS sequence. The TexPREP CS4 course teaches the foundations of computing using ASP [14]. This course emphasized problem solving and knowledge representation through the ASP paradigm. The course met Monday through Friday from 1:00pm to 1:50pm.

Informed consent was obtained from both the participants and their parents prior to the start of class. There were sixteen (N=16) participants for the study. There were seven females and nine males. Participants were students at high school in the Texas Tech area in Lubbock, TX. In terms of rising grades, there were eleven 10$^{th}$ graders, three 11th graders, and two 12th graders.

### 2.2 Data Collection and Procedures

**Surveys.** Participants responded to surveys with open-ended questions at the beginning (pre) and end (post) of the course regarding their experiences with computing and declarative programming. The purpose of the surveys was to provide some background information on participants' CS knowledge and to triangulate the findings from the clinical interviews. The survey refers to declarative programming (DP), rather than LP or ASP even though this study focused on the latter two paradigms. The rationale for referring to the more general DP is because: ASP belongs to the LP paradigm and LP is one paradigm of DP and knowing other DP paradigms (such as functional programming) may give an edge to the students in learning ASP.

The pre-survey asked about participants' general knowledge of computer science and declarative programming. The pre-survey questions were:

- What is computer science?
- Describe what you know about declarative programming.

The post-survey asked about participants' knowledge of computer science and the topics of this course. The post-survey questions specifically asked what participants learned in this course and how they understood the concepts; that is, rather than asking participants directly to explain declarative programming, these questions asked how they perceived declarative programming.

- What is computer science?
- What have you learned, if anything?
- After having taken this course do you feel that you understood the subject of this course and the tasks that were asked of you? Why or why not?

**Clinical Interviews.** Clinical interviews are task-based activities in which the researcher attempts to explore subjects’ understanding and cognitive process through active questioning and probing [19]. During the last week of class, participants were asked to work out one class-based problem while thinking aloud. Since this was a class assignment during a limited time with few researchers, participants only worked on part of the assignment. Clinical interviews were conducted one-on-one by the research team and the participant. Researchers asked probing questions in order to prompt participants to explain and clarify their thought processes. All researchers received the same training from the first and third author on conducting the clinical interviews. Clinical interviews lasted 30 minutes or when the participant completed their tasks—whichever came first. Clinical interviews were video recorded and transcribed.

In the interview task, each student was asked to write a SPARC program to represent the knowledge about family relationships. Some relations, such as *father(X, Y)*, denoting X is the father of Y, and *mother(X, Y)* are given. A program stub including the declarations of these relations is also given. The students are asked to represent the following knowledge: 1) Jon is the father of Matthew, and 2) about the following relations: *grandparent(X, Y)*, *son(X, Y)*, *aunt(X, Y)*, and *descendant(X, Y)*. SPARC is a programming language and system instance of ASP paradigm [20]. It offers a type system to overcome some challenging syntax restrictions in existing ASP systems, such as DLV [21] or Clingo [22] and help discover some programming errors early.

The results of the surveys and interviews are presented in the following sections.

## 3 Survey Results

Survey responses were transcribed and were analyzed by the researchers using an inductive coding approach. The main unit of analysis was a sentence. Longer sentences with representing multiple ideas were sometimes separated for further coding. Similarly, some consecutive sentences were grouped together for analysis when they continued the same line of thought. Through this analysis, researchers found common themes across participants. Responses were grouped by common themes, which are presented below. Participants were given pseudonyms in the presentation of the results in the form of P#.

### 3.1 What LP Students Know about Computer Science

Since this course was the third in the pre-engineering program sequence, participants already had two courses in computing with other science and engineering courses. Most participants' definition of computer science in the pre-survey centered on the creation of a program that does some task with the programming language being the means through which those tasks are completed [e.g., *The science behind how a computer functions and performs tasks* (P1), *Computer science is using different computer "languages" to get computers to perform tasks* (P13), and *Understanding and being able to reproduce the code and programs that make a computer do its task* (P4)]. There was a strong connection between computer science and programming with ten participants mentioning programming or the creation of software in their responses. Some participants acknowledged the importance of computers and technologies in the advancement of society [e.g., *The study on how to develop computers to help the world be an easier and more efficient place* (P10); and *It has become extremely prominent in our daily lives and must always be improved for technology to come* (P11)].

At the end of the course, participants were asked the same question. In the post-survey, only three participants mentioned programming in their definition of computer science. The participants' responses could be categorized as defining computer science as problem solving [e.g., *Computer science is a study that involves mathematics and helps us understand the way a computer thinks* (P3); *Computer science is the study on how to make a computer work/think like a "human"* (P10); *It is how we get computers to solve problems and how we can use them for simplifying, numbers, solutions, or problems themselves* (P11)] and the study of technologies [e.g., *The science behind a computer and how it functions* (P1); *Computer science is the study of how and why computers work* (P16)]. Thus, by the end of the course, the participants' definition of CS had broadened a bit away from just programming; rather, it become more holistic and more on emphasizing problem solving in order to complete tasks.

### 3.2 What LP Students Know about Declarative Programming

In the pre-survey, most participants did not know what DP was; however, many were familiar with the more "traditional" procedural programming and/or had prior programming experience. Eight participants said that they did not know what declarative programming was. Some participants offered an explanation of what declarative programming could be [e.g., *I would guess that it is a branch of computer science that has to do with making statements and receiving some form of output from the statements* (P12); or *I believe that you give a description to of an object then the program runs it to find the item similar to what was described* (P8)] or explanation that they have done programming before [e.g., *In school, I have learned a little about Java and how Java is a declarative language, Other than that, I don't know much else about declarative programming* (P3)]. In such explanations, participants were attempting to make connections to previous programming experiences, even though those were incorrect.

The remaining five participants did not explicitly say they did not know what DP was. Rather, they offered some incorrect or incomplete definitions: *Declarative programming is stating a statement and being able to program that statement. Giving the computer a "command" and programming it to accomplish that particular command* (P5); *It is a program where it does the things you type out* (P10); *Declarative programming is more of an elaboration off of algorithmic programming as it is developed for a broader use compared to Algorithmic programming. It is much more elaborates and is well suited for situations with similarities* (P12); and *It is the type of programming needs to be defined before any output action can be achieved* (P15). Participant 16 was the closest to having a correct answer: *I know that declarative programming works like telling a computer rules for what something is then having the computer rules for what something is then having the computer go find all the somethings [sic] in a set*. Thus, most participants did not have experience with DP though have had some experience in other paradigms, which included object-oriented, procedural, and imperative.

### 3.3 What Students Have Learned in this LP Course

The data showed that most students reported learning how to program with the SPARC language within the context of their lab assignments. Though most did not explicitly articulate their understanding of LP, as they did with procedural-type languages from prior CS experiences, they were able to report how they solved problems using ASP methodologies and tools. Although the participants emphasized learning how to program with SPARC, eleven responses implied that the programming language provided the structure that required students to think about their problems through a deeper, more thorough perspective, which lends itself to declarative programming. P12's response was representative of how learning the programming language helped guide declarative understanding: *I've learned how to reevaluate my thinking process and consider the basis of my thought (so I can program). I've learned how to 'translate' – by writing my goals/rules/knowledge in English, then transferring and modifying it to fit SPARC code.* Similar responses include: *I learned how to look at a problem in a different way* (P2); *I have learned how to sort objects, define relationships with predicates, use multiple variables, use multiple predicates/ rules, use tex*** rules with variables, and use if when for rules to help simplify solutions* (P11); *I have learned a new coding language to go along with the others I know, which is very valuable to me. I have also learned a new way to look [at] relations between objects* (P15); and *New syntax - A new way to specify relations* (P9). These examples show that the rules and structure (i.e., syntax and semantics) that SPARC required and the associated methodologies helped guide participants in their problem solving. Four participants emphasized learning about the syntax and coding style. P8 also mentioned that this was the first course in which she had to type out her programming rather than through a drag-and-drop interface.

### 3.4 How Students Understood the LP Course Topics and Tasks

Eight participants stated that they understood the subject of this course within the confines of the course activities and three were indifferent. They understood that there was more to learn with respect to the programming language [e.g., *I think if given more time and work on the computer, I could use this language in the future* (P3); *The prompt and things that were asked were clear and easy to understand. They were easy to pick up on and gave a general understanding – making it relatively easy to complete a given task* (P5); and *I understand the subject because it was easy. I'm pretty sure that if you wanted to program more complex things, it would be harder* (P13).

Two participants stated they were able to understand the subject of the course, but were unsure how it was applicable in a real world or everyday setting: *I understand how, but not why. So, I don't understand the purpose of this type of program and how it will be beneficial to the world, other than artificial intelligence, but I feel as though this program requires more to get to that point* (P9) and *I do feel I understood the subject of the course as I did actually learn something. However, I did not always understand exactly what I was supposed to do and I was occasionally completely lost* (P4). Therefore, most participants knew that there was more to learn about LP beyond the contexts with which they were presented. The two participants (P10 and P14) that stated they did not understand the course topics mainly due to time management in the course. P10 felt that the course was too repetitive with the simple topics and was not challenging enough. Similarly, P14 felt the course was also repetitive with instructor teaching topics already found in their notes.

## 4 Interview Results

A grounded theory methodology [16] was adopted to generate a theory that explained how students understood and approached computing through ASP activities based on the interview data. This qualitative approach generates a theory or explanation through systematic analysis of data [16]. Grounded theory goes beyond describing and categorizing by finding the processes and relationships in the phenomenon being studied [16]. This approach was conducted on the clinical interview data and served as the main analysis for this study. The unit of analysis was an utterance, which consisted of a complete thought or phrases spoken by the participant. In cases where utterances were rich with many potential units, they were broken down into smaller utterances. Utterances made by the researcher provided context to those made by the participant, but were not analyzed. There were 1,452 utterances used for analysis.

The first author analyzed the data through an open-coding process [16] coding each utterance with a descriptive label. This process required several iterations to ensure that coding was consistent across the entire dataset. Then, the third author coded the data using the set of labels created by the first author. Through that process, the researchers debated the labels and coding of the utterances, which led to several iterations of re-coding the data and revision of the initial labels. Memos were kept during this analysis process regarding the rationale behind labels and the formation of coding [16]. Since the third author was the expert in the field of ASP and DP and taught the course, his coding was used for the remaining analysis.

The next step was creating categories that represented larger constructs that was happening the data. Axial coding analyzed the way these categories were related based on their properties [16] and formed larger categories. In the case of this study, many of the merged labels were found to relate to each other as well with infrequent labels. The five major categories were: abstraction, representation, reasoning, revision, and procedural. The first author led the first stage of axial coding which led to these five major categories. Then, these were also debated and discussed by the researchers. The surveys, particularly the post-surveys, were used to guide the axial coding and the creation of the categories. The definitions of the categories also went through several revisions. The next sections define each category and give example quotes.

### 4.1 Abstraction

This category describes instances where the participants understand the problem space in more general or abstract terms and are thinking on a higher level or in real world terms and within the context of the code they are writing. Participants are generally describing relationships with respect to these real-world terms and/or integrating their prior knowledge of the problem space.

Example quotes included: *(P8) The descendant part because the descendants can be anything that's…that was before that person, well before X in this case.; (P10) So, father of person and person, ok, parent, sister, same parent, gender; (P12) I am thinking of like a family tree I guess sort of and so I am thinking of X being or I am thinking of Y being so I am thinking of Y being a person, whose descendants we are looking at…*

### 4.2 Representation.

This category refers to the process in which the participants are trying to understand the problem space in coding terms; that is, how their abstraction or understanding of the problem translates into the code. This category includes references to both the conceptual and coding as well as the transition process between the two.

Example quotes included: *(P1) You can't say, X is the parent of Y if X, if Y is the son, because it's flipped. I hope I am explaining it right. So then I'm saying X is the son of Y if Y is the parent of X, and I'm putting gender because if you are a son you have to be male. And if X is a male. Yes, Ok. All right I am going to move on to aunt.; (P14) A descendant is like it's the same as a child it's umm like the child of two parents. That is their descendant that's their heir so that is what I am trying to say but I don't know how to say it.; (P15) I put X is the son of Y. I put that in the predicates. This is it. The son is a person. And I know that we have the example here. I can put Matthew is the son of Jon…*

### 4.3 Reasoning

This category represents the problem solving process. It describes instances in which participants were actively thinking about the problem, apply problem-solving strategies associated with ASP, and adapting existing code to solve the current problem. Reasoning was selected for the category name as that the most frequent related label.

Example quotes included: *(P2) So grandchild Isaac of George, grandchild Joseph, George, grandchild Susie, George. Grandchild, those are the children, so, grandchild, grandchild, X is a grandchild of Y, if Y is grandfather, if Y is the grandfather of X and X is the, do we define gender? Yes, gender of X is male, because grandfather has to be a male; (P5) I have like separate ones like I have sister and I have like aunt and I have parent; (P11) Nephew, so then you have Isaac, and actually you can create a rule...*

### 4.4 Revision

Revision encapsulates a lot of the overall debugging process, but emphasizes the importance of participants asking questions in SPARC to see if their code was correct. This category includes the participants' process of questioning to see if they completed the tasks correctly, which was done through querying to see if their solutions were right.

Example quotes included: *(P1): So I'll write aunt and then I'll write Rose comma = Susie, and then I'll have to add a question mark, and then have to press execute; (P14): Ok and now I need to write queries to see if my program works; (P12): I am going to ask who are descendants of George because George is like the patriarch of the family…*

### 4.5 Procedural

This category represents a large part of the data in which participants were either restating or interpreting facts or instructions. In some cases, participants were asking if they have the correct interpretation of the assignment instructions. In this category, participants frequently mentioned specific rules and strategies associated with declarative programing. Often, these utterances were co-labeled with other more descriptive categories such as representation and reasoning.

Example quotes included: *(P1): Whenever I am talking about the son I forgot to include that it has- and he has to be male, and that X has to be male.; (P8): So now I am at son. Oh forgot to add the periods. Ok. Got it, got it, and got it. Now I need it aunt.; (P16): Jon is the father of Matthew, so that rule that I wrote works.*

### 4.6 Order of Category Occurrences

As part of axial coding, there was an analysis of when these categories occurred during the clinical interview to learn more about the categories themselves and how they related to the other categories. A color-coded graph plotted out the occurrence of each category in the order they appear with each participant. Plots varied in length due to the variance in number of coded utterances. Figure 1 shows an example of this graph. When an utterance had more than one category associated with it, plots are stacked.

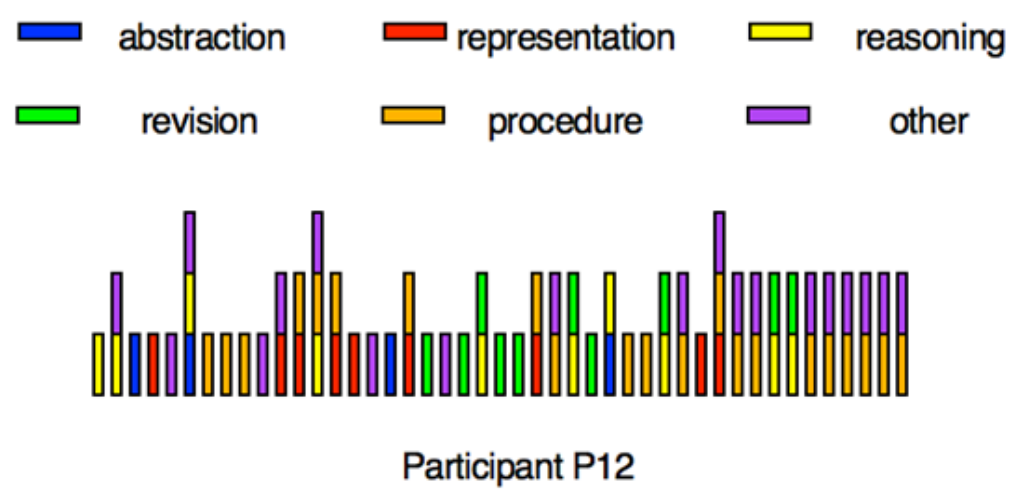

**Fig. 1.** Sample of Order of Category Occurrence

A visual analysis of all participants' graphs was conducted to describe when each category occurred and the relationships between those occurrences. Across all participants, interviewers had them all read the instructions once at the beginning, but these utterances were not coded. Since participants were asked to work on their assignments, some participants had already started on them prior to the interview. Thus, those graphs immediately started with non-procedural labels. The visual analysis found the following:

**Revision.** Ten participants were engaged in revision during the last half the interview, which was expected as they spent the first half identifying the objects, relationships, knowledge, and coding before testing. P10 did not get as far during the interview to be engaged in revision. Other participants had already started their assignments before the interview, so they were quicker to be in revisions.

**Abstraction.** Evidence of abstraction mostly came from the beginning to the middle for most participants. For several participants, it occurred alongside procedural (7 participants) as well as with representation (6 participants). Three participants had abstraction after revision, which may suggest that the questioning process may support part of abstraction abilities. Another three participants showed no evidence of abstraction during the interview. However, the interviewer reported that P2 did not get a chance to test their program and P3 had technical issues with the laptop. P13 was reported to feel confident about what they were doing, and started revision early.

**Reasoning.** Reasoning came throughout interview for most participants, similar to the procedural utterances, which may suggest that it is an important recurring process. Reasoning occurred with and between instances of revision (P1, P16) and representation (P2, P13), abstraction (P14, P15). Thus, reasoning is also intertwined with other parts of the ASP process. There was no evidence reasoning from P6; however, there were not many coded utterances from that participant in general.

**Representation.** Similar to reasoning and procedural, representation also happened throughout the interview. For the most part, representation started before clusters of revision for 11 participants. However, for three of these participants, there were more revision afterward. For P13, representation was intertwined into revision throughout the interview. Most of P14's interview consisted of utterances labeled as representation as he was mostly listing out what had to be represented in the knowledge base.

**Procedural.** All participants had instances of procedural utterances and distributed throughout the interview as well as simultaneously with other codes. An interesting observation was that there were procedural moments throughout the interview, but not just at the beginning. That is, participants kept going back to the instructions and/or the facts presented in the assignment.

Although the procedural category did not seem to provide much insight into how the participants approached computing and computer science, at first. After all, these utterances were of participants reading and re-reading the rules and asking if they understood the instructions and information correctly. Procedural utterances were made throughout the entire overall computing process. In selective coding where analysis examines a category that could be connected to the other categories [16], procedural was found as that central theme. Identifying objects and relations in the problem, through English definitions, by carefully reading the problem description and integrating one's common sense knowledge was a major component of the explicit methodology for solving LP problems. Thus, there was deep interaction between the procedural and other processes. Students had also mentioned in their post surveys that they had learned this methodology to solve the problems. Although that may not be surprising, it showed that they were able to adopt those strategies, which in turn, led to more reasoning and representation processes.

### 4.7 Model of How Participants Approach LP

Based on the grounded theory analysis, supported by the visual analysis of ordering of categories, model was constructed that may explain how participants approach computing through LP as based on the data (see Figure 2).

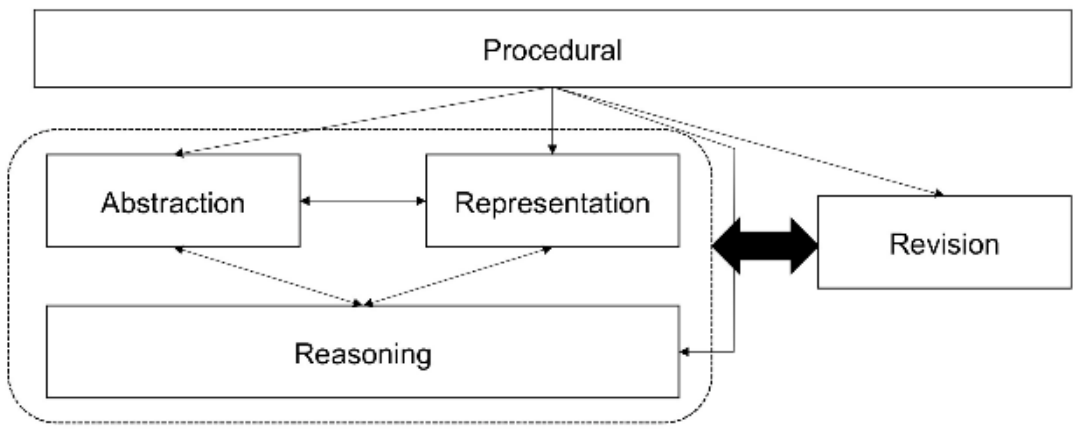

**Fig. 2.** Model of How Participants Approach LP

This figure highlights the foundational importance of the following the strategies of declarative programming. In the case of Answer Set Programming, it provided an explicit methodology, which guided students towards abstracting the concepts, representing it in code, and reasoning in their problem solving. Abstraction, representation, and reasoning often happened together and are grouped separately from revision, which mostly came toward the end of the task. Similarly, abstraction and representation happened more sequentially with abstraction coming before representation. Reasoning was observed throughout the interview, which is indicated by a larger rectangle.

The model showed the importance of procedures; in this case, students were able to follow the explicit methodology in LP. The students also demonstrated the general problem solving skill of iterative refinement. There are two distinguishable big loop components: knowledge specification and revision. Knowledge specification is again an iterative refinement using interleaving steps: abstraction, representation and reasoning. As an example, when students try to specify their knowledge about aunt, they try to figure out the objects (individual persons) and relations (e.g., who is whose sister, or who is whose parent) and then try to define their knowledge about aunt using these objects and relations. Participants tried to make sure the definition captures their intended meaning, which requires them to reason with and understand the relations better. In fact, this model reflects the intended skills most researchers would like the students to obtain in an introductory course: the skills should be general but important in computing (and beyond). The skills include the procedure (a general methodology for problem solving), iterative refinement in problem solving, programming (revision in our category name), abstraction, rigorous high level (logic) representation of knowledge and (logic based) reasoning with the knowledge.

LP can be thought of as introducing an explicit problem solving methodology: problem understanding and precise representation and reasoning with knowledge. This methodology is almost universal in problem solving. It will also particularly be useful as LP lays the foundation for problem understanding and knowledge representation.

## 5 Discussions

The students had done very well in their lab assignments and final interview tasks, which, together with the pre and post surveys, which shows that LP is easy to learn [8].

The main reason is the simple syntax and intuitive/natural semantics of ASP, which allow the students to focus less on language specific feature and more on the problem solving skills. The clinical interview data shows further that the students were able to learn and apply intensively the important concepts underlying computing. The model (Figure 2) shows that they were able to apply explicit LP methodologies in problem solving and iterative refinement in both big steps (knowledge specification and revision) and small steps (abstraction, representation and reasoning), when solving the problems. Participants applied abstraction, representation and reasoning heavily, in an interwoven manner, in understanding, extracting and defining the knowledge needed in solving a given problem. Once they identified/defined the knowledge needed, they are able to carry out the standard programming tasks (labeled as “revision” in our categories) from coding to debugging the program. The model generated showed that students were able to engage in computing skills, such as abstraction, problem solving (representation), and debugging (revision). Indeed, they were able to explain computing concepts within the context of this course and tasks. Students were able to operate within the LP paradigm driven by the explicit methodology for LP; however, they may not immediately see how it is applicable outside the context of the assignments. They understood the LP concepts and processes and were successful in completing their assignments. More work needs to be done to make connections to how they see CS in the real world through LP. Students understood the problem solving nature of CS and believed LP was one way of solving problems.

### 5.1 Limitations

Participants were above average students and were pre-screened for this program, and had some experience with Scratch and Alice. However, participants stated no experience with LP or DP. Though every student initially volunteered for this study, there were still some participants who stopped attending or did not complete an assignment. Lastly, this course met for four weeks, which provided limited exposure to LP as compared to regular K-12 or university courses. Future studies on LP in introductory CS courses could be conducted on full semester courses rather than short courses.

## 6 Conclusions

This paper asserts that CS educators should take a closer look at using LP during the introductory courses. As suggested in [12], future computer scientists should be equipped with foundational programming language principles involving logic and formal specification to design and implement complex software systems needed by the society. Our results show that the students were able to focus on the key concepts of computing including abstraction, representation and reasoning when solving problems. The simple syntax and intuitive semantics of ASP allows them to put less attention on the language specific features. These findings support LP as a viable option to teach an introductory CS course using ASP.

**Acknowledgments.** The authors acknowledge Cynthia Perez and Rocky Upchurch for their contributions to this project, and thank Michael Gelfond for sharing his teaching materials. This work is partially supported by NSF grant CNS-1359359.